\documentclass[twocolumn,showpacs,prl,superscriptaddress,floatfix]{revtex4}

\usepackage{color} 
\usepackage{tabularx} 
\usepackage{epsfig}
\usepackage{amsmath} 
\usepackage{amssymb} 
\usepackage{graphicx}

\begin{document}

\title{Ultrametricity and clustering of states in spin 
glasses: A one-dimensional view}

\author{Helmut G.~Katzgraber} 
\affiliation {Theoretische Physik, ETH Zurich, CH-8093 Zurich, Switzerland}
\affiliation {Department of Physics, Texas A\&M University, College Station,
Texas 77843-4242, USA}

\author{Alexander K. Hartmann}
\affiliation {Institut f\"ur Physik, Universit\"at Oldenburg, D-26111
Oldenburg, Germany }

\begin{abstract}

We present results from Monte Carlo simulations to test for
ultrametricity and clustering properties in spin-glass models. By using
a one-dimensional Ising spin glass with random power-law interactions
where the universality class of the model can be tuned by changing
the power-law exponent, we find signatures of ultrametric behavior
both in the mean-field and non-mean-field universality classes for
large linear system sizes. Furthermore, we confirm the existence
of nontrivial connected components in phase space via a clustering
analysis of configurations.

\end{abstract}


\maketitle

An ultrametric (UM) space \cite{rammal:86} is a special kind of
metric space in which the triangle inequality $d_{\alpha\gamma}\le
d_{\alpha\beta} + d_{\beta\gamma}$ [$d_{\alpha\beta}$ represents the
distance between two points $\alpha$ and $\beta$] is replaced by a
stronger condition where $d_{\alpha\gamma}\le \max\{d_{\alpha\beta},
d_{\beta\gamma}\}$, i.e., the two longer distances must be equal and
the states thus lie on an isosceles triangle. The concept appears in
many branches of science, such as p-adic numbers, linguistics, as well
as taxonomy of animal species.  It is also an intrinsic property of
Parisi's mean-field solution \cite{parisi:79,mezard:84-ea,binder:86} of
the Sherrington-Kirkpatrick (SK) \cite{sherrington:75} infinite-range
spin glass.  Hence, in general, the nature of the spin-glass state
\cite{binder:86,mezard:87} can be analyzed via clustering and
ultrametricity-probing methods.

The nature of the spin-glass state is controversial and it is
unclear if the mean-field replica symmetry breaking (RSB) picture
\cite{parisi:79}, the droplet picture \cite{bray:86,fisher:86},
or an intermediate phenomenological scenario dubbed as TNT
\cite{krzakala:00,palassini:00} (for ``trivial--nontrivial'') describes
the nature of the spin-glass state best. One way to settle the
applicability of the RSB picture to short-range (SR) spin glasses is by
testing if the phase space is UM.  Unfortunately, the existence of an
UM phase structure for SR spin glasses is controversial, mainly because
only small linear system sizes have been accessible so far.  Recent
results \cite{hed:03} suggest that SR systems are not UM, whereas other
opinions exist \cite{franz:00,contucci:07-ea,contucci:08-ea,joerg:08c}.
Thus it is of paramount importance to test if SR spin glasses have
an UM phase space.

In this work we approach the problem from a different angle: First,
we use a one-dimensional (1D) Ising spin-glass with power-law
interactions.  The model has the advantage that large linear system
sizes can be studied.  Furthermore, by tuning the exponent of the
power law, the universality class of the model can be tuned between a
mean-field and a non-mean-field universality class. This allows us to
test our analysis method on the mean-field SK model and then apply it
to regions of phase space where the system is not mean-field like.
We perform a clustering analysis of the data similar to the work
of Hed {\em et al.}~\cite{hed:03} to obtain nontrivial triangles in
phase space and introduce a novel correlator which allows us to see
an UM signature for low temperatures and delivers no signal for high
temperatures.  Furthermore, we use a clustering analysis to search
for connected components in phase space.  The proposed method can be
applied to any field of science to test for an UM structure of phase
space, thus making the method generally applicable.

Our results for low temperatures show that for this model the phase
space has an UM signature and exhibits many phase-space components,
the number growing with system size in the mean-field as well as
non-mean-field case. This suggests that for large enough system sizes
SR spin glasses at low enough temperatures might have an UM
phase space structure.

\begin{figure}

\includegraphics[width=0.85\columnwidth]{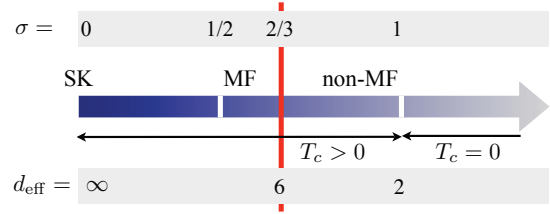}

\vspace*{-0.3cm}

\caption{(Color online)
Sketch of the phase diagram of the 1D Ising chain with random power-law
interactions.  For $\sigma \le 1/2$ we expect SK-like infinite-range
behaviour. For $1/2 < \sigma \le 2/3$ we have mean-field (MF) behaviour
corresponding to an effective space dimension $d_{\rm eff} \ge 6$,
whereas for $2/3 < \sigma \le 1$ we have a long-range (non-MF)
spin glass with a ordering temperature $T_{\rm c} > 0$. Close to
$\sigma = 2/3$ (vertical red line) $d_{\rm eff} \approx 2/(2\sigma -
1)$ \cite{binder:86}.  For $ \sigma \ge 1$, $T_{\rm c} = 0$.
}
\label{fig:pd}
\end{figure}

\paragraph*{Model.---}
\label{sec:model}

The Hamiltonian of the 1D Ising chain with long-range
power-law interactions  \cite{kotliar:83,katzgraber:03}
is given by
\begin{equation}
{\mathcal H} = - \sum_{i<j} J_{ij} S_i S_j
\;\;\;\;\;\;\;\;\;\;\;\;
J_{ij}= c({\sigma}) \frac{\epsilon_{ij}}{{{r_{ij}}^\sigma}} \, ,
\label{eq:model}
\end{equation}
where $S_i \in\{\pm 1\}$ are Ising spins and the sum ranges over all
spins in the system. The $L$ spins are placed on a ring and $r_{ij}
= (L/\pi)\sin(\pi |i - j|/L)$ is the distance between the spins.
$\epsilon_{ij}$ are Normal random couplings. The constant $c(\sigma)$
is chosen such that the mean-field transition temperature to a
spin-glass phase is $T_c^{\rm MF} = 1$ \cite{katzgraber:03}.

The model has a very rich phase diagram when the exponent $\sigma$ is
tuned \cite{katzgraber:03}: Both the universality class and the range
of the interactions of the model can be continuously tuned by changing
the power-law exponent, see Fig.~\ref{fig:pd}.  In this work we study
the SK model [$\sigma = 0$, $T_c = 1$] to test our analysis protocol,
as well as the 1D chain for $\sigma = 0.75$ [$T_c \sim 0.69$] and
$0.85$ [$T_c \sim 0.49$] \cite{katzgraber:05c}; both corresponding to
the non-mean-field regime. We choose two values of $\sigma$ to be able
to discern any trends when the effective dimensionality is reduced.

\paragraph*{Numerical details.---}
\label{sec:meas}

We generate spin-glass configurations by first equilibrating the
system at $T \approx 0.4 T_c$ using the exchange Monte Carlo method
\cite{hukushima:96}, i.e., $T = 0.4$ for the SK model, $0.27$ for
$\sigma = 0.75$ and $0.20$ for $\sigma = 0.85$.  Once the system
is in thermal equilibrium we record states ensuring that these are
well separated in the Markov process and thus not correlated by
measuring autocorrelation times.  In practice, if we equilibrate the
system for $\tau_{\rm eq}$ Monte Carlo sweeps, we generate for each
disorder realization $10^3$ states separated by $\tau_{\rm eq}/10$
Monte Carlo sweeps.  We test equilibration using the test presented
in Ref.~\cite{katzgraber:05c}; see Table \ref{tab:simparams} for
simulation details.

\begin{table}
\caption{
Simulation parameters for the 1D chain and different power-law
exponents $\sigma$. $L$ is the system size, $N_{\rm sa}$ is the
number of disorder realizations, $\tau_{\rm eq}$ is the number of
equilibration sweeps, $T_{\rm min}$ is the lowest temperature and
$N_{\rm r}$ the number temperatures used in the exchange Monte
Carlo method.  \label{tab:simparams}}
{\footnotesize
\begin{tabular*}{\columnwidth}{@{\extracolsep{\fill}} l@{\hspace{-0.75em}}l@{\hspace{-0.75em}}l@{\hspace{-0.75em}}l r r r c c }
\hline
\hline
$\sigma$ & & & &   $L$  &  $N_{\rm sa}$  & $\tau_{\rm eq}$ & $T_{\rm min}$ & $N_{\rm r}$\\
\hline
$0.00$ & $0.75$ & $0.85$ &        &   $32$ & $4\,000$ &  $10\,000$ & $0.20$ & $20$ \\
$0.00$ & $0.75$ & $0.85$ & $4.00$ &   $64$ & $4\,000$ &  $10\,000$ & $0.20$ & $20$ \\
$0.00$ & $0.75$ & $0.85$ & $4.00$ &  $128$ & $4\,000$ &  $10\,000$ & $0.20$ & $20$ \\
$0.00$ & $0.75$ & $0.85$ & $4.00$ &  $256$ & $4\,000$ &  $65\,000$ & $0.20$ & $20$ \\
$0.00$ & $0.75$ &        &        &  $512$ & $2\,000$ & $200\,000$ & $0.20$ & $20$ \\
       &        & $0.85$ &        &  $512$ & $2\,000$ & $650\,000$ & $0.20$ & $20$ \\
$0.00$ &        &        &        & $\,1024$& $1\,000$ &  $32\,000$ & $0.40$ & $26$ \\
\hline
\hline
\end{tabular*}
}
\end{table}

\paragraph*{Analysis details.---} 
\label{sec:analysis}

\begin{figure}

\includegraphics[width=6.3cm,angle=270]{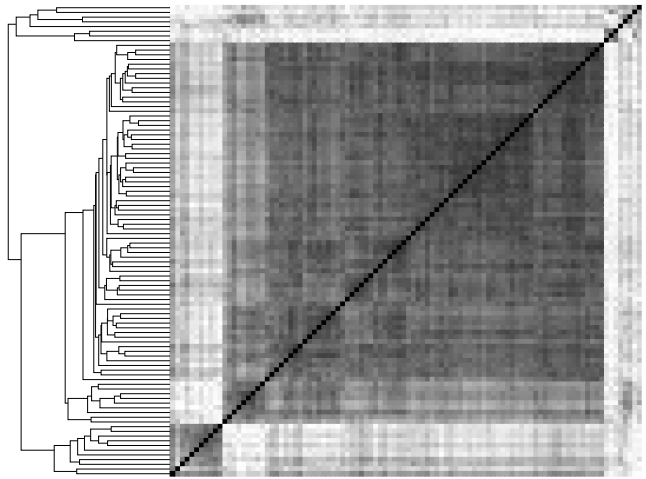}
\caption{
A dendrogram obtained by clustering 100 configurations (see text)
for a sample system with $\sigma=0.0$ and $L = 512$ at $T=0.4$
together with the matrix $d_{\alpha\beta}$ shown in grey scale
(distance 0 is black). The order of the states is given by the leaves
of the dendrogram (figure rotated clockwise by $90^\circ$).
\label{fig:matrix}}
\end{figure}

We use an approach closely related to the one used by Hed
{\em et al.}~\cite{hed:03}. $M = 10^3$ equilibrium states at $T
\approx 0.4 T_c$---to probe deep within the spin-glass phase---are
sorted using the average-linkage agglomerative clustering algorithm
\cite{jain:88}: Distances are measured in terms of the hamming distance
$d_{\alpha\beta} =(1-|q_{\alpha\beta}|)$, where $q_{\alpha\beta}
= N^{-1}\sum_iS_i^{\alpha}S_i^{\beta}$ is the spin overlap between
states $\{S^\alpha\}$ and $\{S^\beta\}$.  The clustering procedure
starts with $M$ clusters containing each exactly one state. Distances
between clusters are introduced, which are initially equal to the
distances between the corresponding states. Iteratively the two
closest clusters $C_a$ and $C_b$ are merged into one cluster $C_d$,
reducing the number of clusters by one. The distances of the  new
cluster $C_d$ to the other remaining clusters have to be calculated:
The distance  between two clusters is the average distance between
all pairs of members of the clusters. The procedure is iterated until
one cluster remains.  The sequence of mergers can be displayed by a
tree, referred to as a {\em dendrogram}.  The root of the dendrogram
corresponds to the last cluster, while the leafs correspond to the
initial states, see Fig.~\ref{fig:matrix}. Furthermore, we also
show in Fig.~\ref{fig:matrix} the distance matrix $d_{\alpha\beta}$
having ordered the states according to the leaves of the dendrogram.
The matrix elements are encoded in gray scale (black corresponds to
zero distance). The complex phase-space structure is clearly visible:
The matrix has a block-diagonal form, the blocks again being subdivided
in a block-diagonal structure.

To analyze the matrix quantitatively for ultrametricity, we
randomly select three states from different branches of the tree
\cite{remark:tree} and sort the distances: $d_{\rm max} \geq
d_{\rm med} \geq d_{\rm min}$. We compute the correlator
\begin{equation} 
K = (d_{\rm max} - d_{\rm med})/\varrho(d) ,
\label{eq:K}
\end{equation}

where $\varrho(d)$ is the width of the distribution of distances.
Note that the definition of $K$ in Eq.~(\ref{eq:K}) differs from
the definition used in Ref.~\cite{hed:03} where the normalization is
performed with $d_{\rm min}$.  Our choice ensures that any apparent
change of an UM measure is scaled out, which is just caused by a
width change of the distance distribution.  The definition used
in Ref.~\cite{hed:03} can {\em only} tell if there is {\em no}
ultrametricity, i.e., a random bit string will also show an UM
response.  The definition in Eq.~(\ref{eq:K}) alleviates this problem:
For $T > T_c$ (or a random bit string) there is no UM signature in
$K$, whereas for $T \ll T_c$ we see a clear UM response for the SK
model. Thus we are able to discern between ``trivial ultrametricity,''
which occurs from equilateral triangles at $T > T_c$, and a true UM
phase-space structure. If the phase space is UM then we expect $d_{\rm
max} = d_{\rm med}$ for $L \rightarrow \infty$. Thus $P(K) \rightarrow
\delta(K = 0)$ for $L \rightarrow \infty$ \cite{comment:eytan}.
We have also computed $P(K)$ for a Migdal-Kadanoff spin glass
\cite{comment:thomas} finding no UM signal for systems up to $N =
149798$ spins, as one would expect.

We also analyze the connected components in phase space (visible
in the distance matrices $d_{\alpha\beta}$) by extending the
approach of Kelley {\em et al.}~\cite{kelley:96-ea}.  During the
$i$'th iteration of the clustering algorithm one encounters $M(i) =
M - i$ clusters.  Thus the goal is to find the number of clusters
which represents the data best corresponding to the highest-level
blocks in the ordered $d_{\alpha\beta}$ matrix.  To obtain a better
resolution at the scale of small distances, we use a logarithmic
scale $\tilde{d}_{\alpha\beta} \sim 1 - \log d_{\alpha\beta}$,
normalized to values $[0,1)$ \cite{comment:log}.  To measure the
component property of the configuration space, we calculate for
each cluster $\Gamma=\{\alpha_i\}$ obtained during the algorithm
the average distance within the cluster (``spread'') $sp_\Gamma=
2\sum_{\alpha\neq \beta \in \Gamma} \tilde{d}_{\alpha\beta}
/|\Gamma|(|\Gamma|-1)$.  Here $|\Gamma|$ is the number of states in
the cluster $\Gamma$.  Then, for each iteration $i$, the average
spread $\overline{sp}_i$ among the $M(i)$ clusters is calculated.
Once the clustering analysis is completed, all $M$ average spread
values are normalized to lie in the interval $[1,M-1]$, resulting
in $\overline{sp}_i^{\rm norm}$.  For each realization the minimum
$M_{\min}$ of $\overline{sp}_i^{\rm norm}+\gamma M(i)$ as a function
of $M(i)$ is determined, where $\gamma$ is a sensitivity parameter
(the method of Ref.~\cite{kelley:96-ea} corresponds to $\gamma = 1$).
Then $n_{\rm C}=M_{\min}$ is the number of phase-space components.
Note that the larger $\gamma$ is, the fewer components are found. Since
a paramagnet should exhibit only one component, we determine for
each system size $L$ $\gamma(L)$ such that for $M=10^3$ random bit
strings ($T = \infty$), averaged over $10^2$ runs, on average $1.01$
components are obtained \cite{comment:clusters}.

\paragraph*{Results.---}
\label{sec:results}

\begin{figure}

\includegraphics[width=0.9\columnwidth]{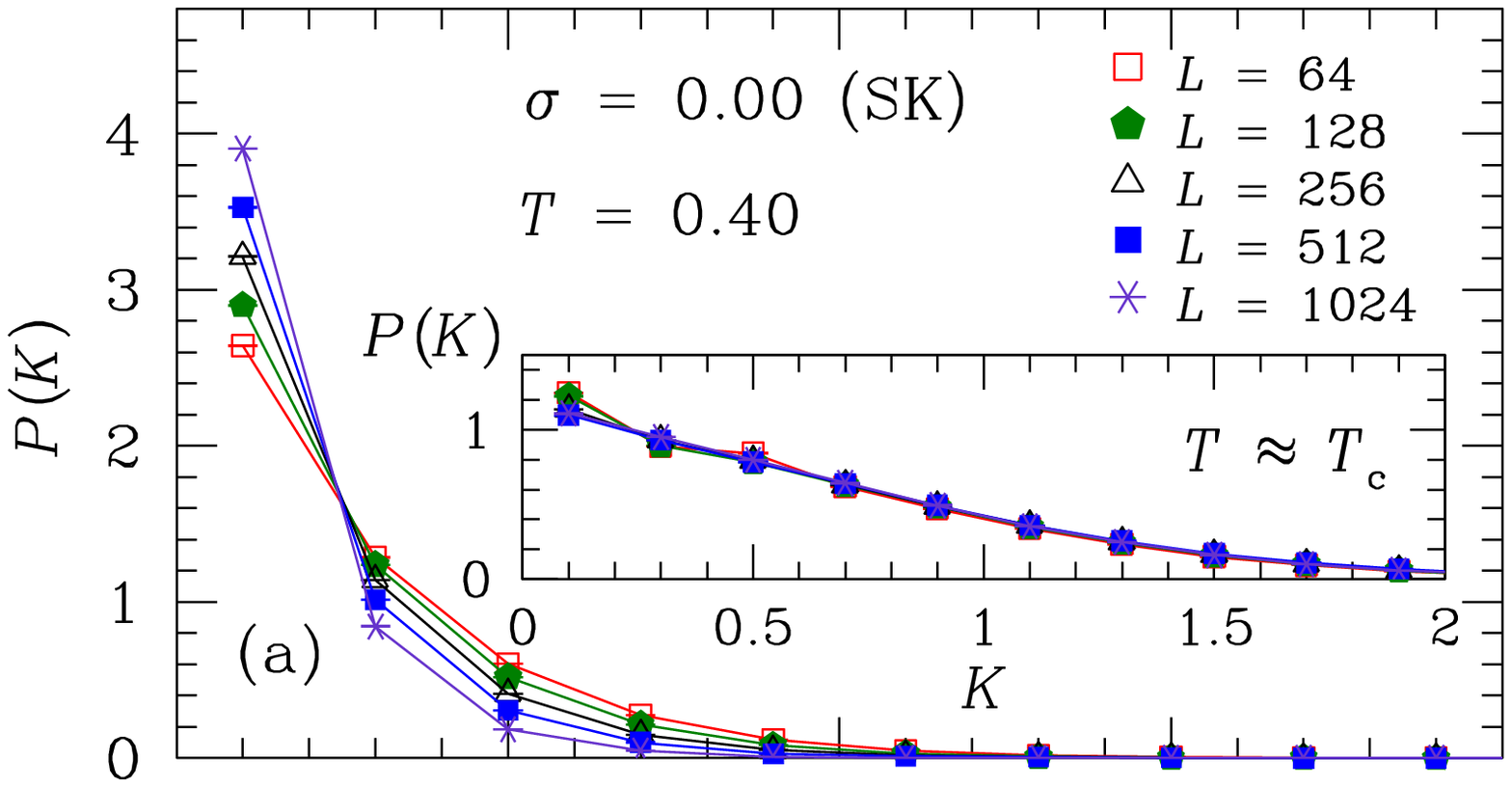}

\vspace*{-0.3cm}

\includegraphics[width=0.9\columnwidth]{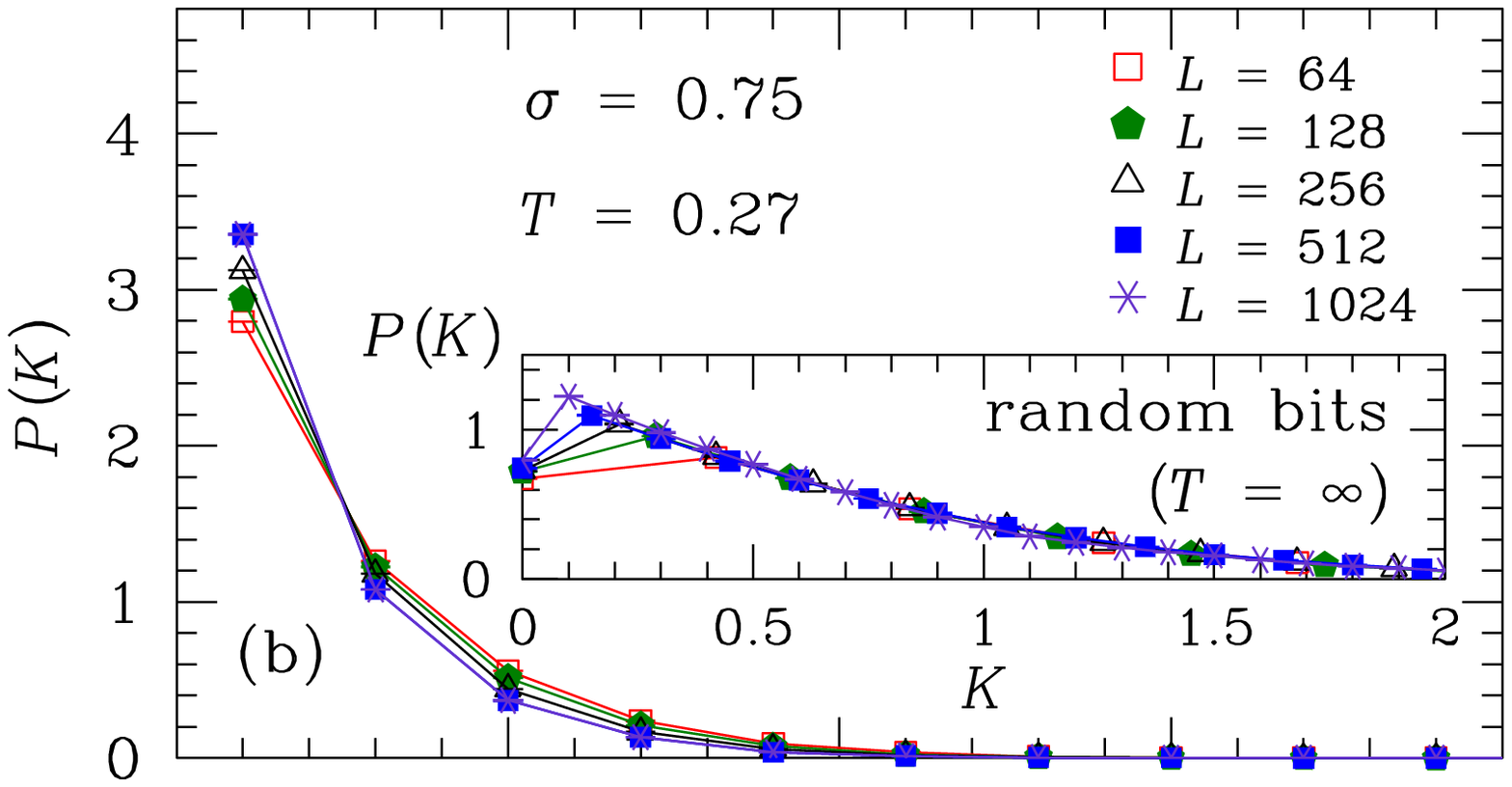}

\vspace*{-0.3cm}

\includegraphics[width=0.9\columnwidth]{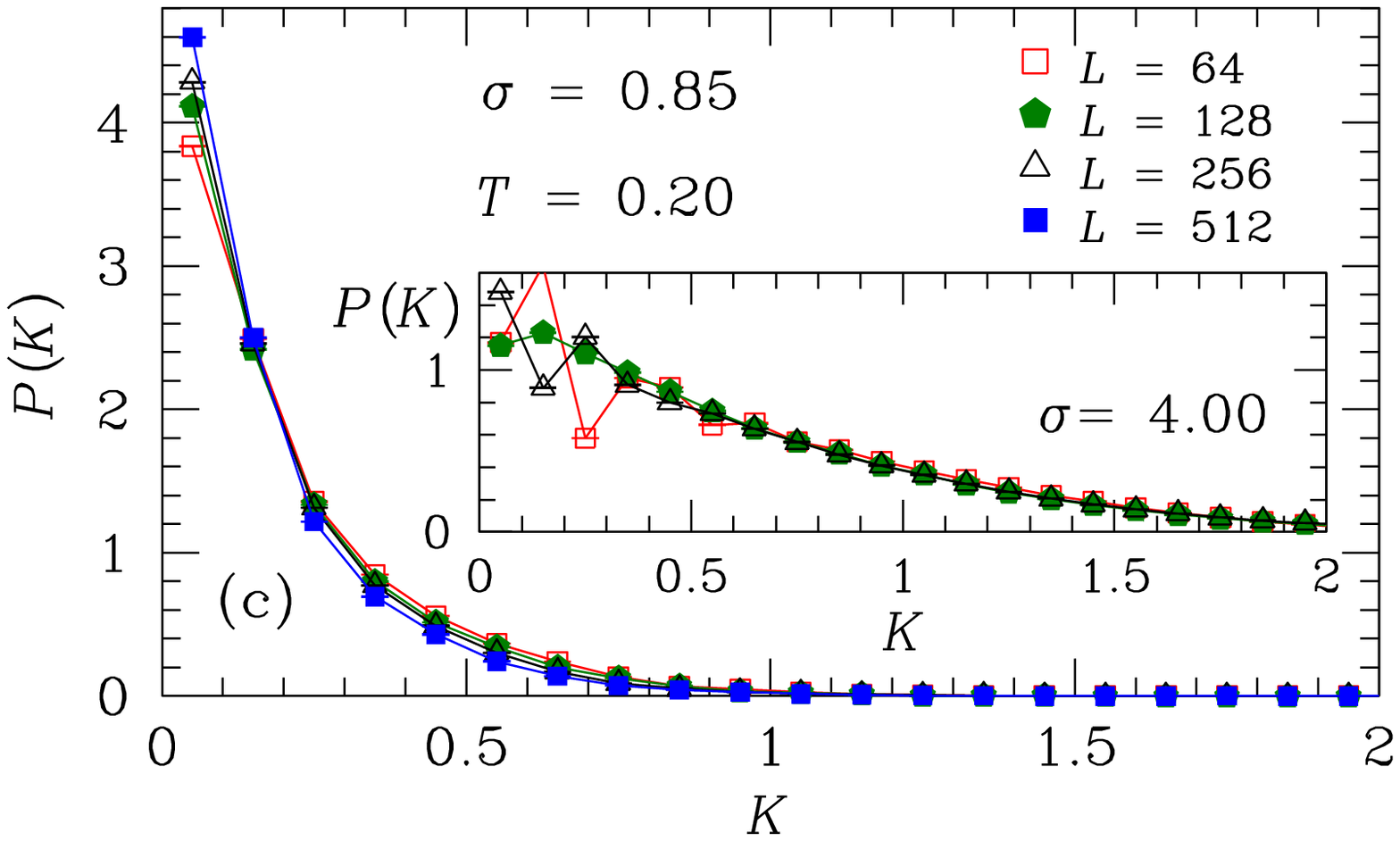}

\vspace*{-0.6cm}

\caption{(Color online) 
Distribution $P(K)$ for different system sizes (all panels have the
same horizontal scale). (a) Data for the SK model. The distribution
diverges for $K \to 0$ thus signaling an UM phase structure. Inset:
For $T \sim T_c$ no divergence is visible.  (b) Data for the 1D
chain for $\sigma = 0.75$ (non-mean-field universality class). The
distribution diverges. Inset: For random bit strings the distribution
shows no divergence.  (c) $\sigma = 0.85$
(non-mean-field universality class). The distribution diverges.
Inset: For $\sigma = 4$ (corresponding to a system with $d_{\rm eff}
\le 2$ the distribution shows no divergence.
}
\label{fig:pk}
\end{figure}

\begin{figure}

\includegraphics[width=0.9\columnwidth]{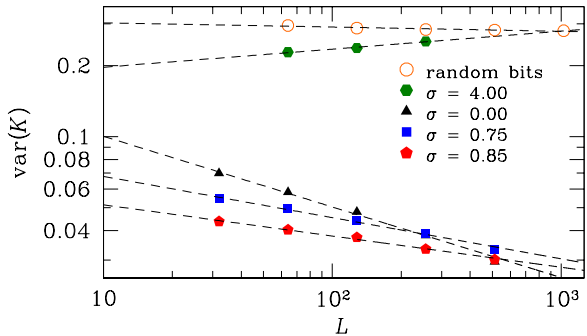}

\vspace*{-0.6cm}

\caption{(Color online)
Variance of $P(K)$ as a function of $L$ for different $\sigma$. The
data are well fit to a power-law decay $\sim b/L^c$ (dashed lines) with
$c = 0.39(10)$, $0.16(2)$, and $0.13(1)$ for $\sigma=0.0$, $0.75$, and
$0.85$, respectively, suggesting a divergence for $K\to 0$ for $\sigma
\lesssim 1$. We have also computed the ``fraction of UM instances,''
(those which exhibit $\int_0^{0.5}P_{\rm instance}(K)\,dK \ge 0.5$,
not shown). For larger system sizes, this fraction grows with the
system size for $\sigma<1$ values.  Hence the results for $2/3< \sigma <
1$ are not due to rare strongly-ultrametric instances.
}
\label{fig:variance}
\end{figure}

In Fig.~\ref{fig:pk} the distribution $P(K)$ is shown for $\sigma =
0$ (SK model), $0.75$ (non-mean-field), $0.85$ (non-mean-field) for
$T \approx 0.4 T_c$.  In all three cases, $P(K)$ seems to converge
to a delta function for $L \to \infty$.  This is clearly visible
when looking at the variance of the distribution which decays
with a power-law of the system size (see Fig.~\ref{fig:variance}).
Note that $P(K)$ does not change with system size close to $T_c$
[inset to Fig.~\ref{fig:pk}(a)].  A similar lack of divergence has
also been found for simulations for $\sigma = 4.0$ in the SR 
universality class [inset to Fig.~\ref{fig:pk}(c)]. For random
bit strings [inset to Fig.~\ref{fig:pk}(b)] the correlator also
shows no sign of UM.  Therefore, the correlator [Eq.~(\ref{eq:K})]
can clearly distinguish between ``trivial'' ultrametricity---which
is due to equilateral triangles---and ultrametricity created by
a complex energy landscape. We have also performed an equivalent
analysis by replacing the spin overlap $q_{\alpha\beta}$ by the
link overlap $q_{\alpha\beta}^l = N_{\rm bonds}^{-1}\sum_{i<j}
S_i^{\alpha}S_j^{\alpha}S_i^{\beta}S_j^{\beta}$.  In this case, for
all values of $\sigma$, $P(K)$ does not converge to a delta function.
This is to be expected, since a different approach is needed
\cite{contucci:07-ea} to obtain evidence for UM using $q_{\alpha\beta}^l$.

\begin{figure}

\includegraphics[width=0.9\columnwidth]{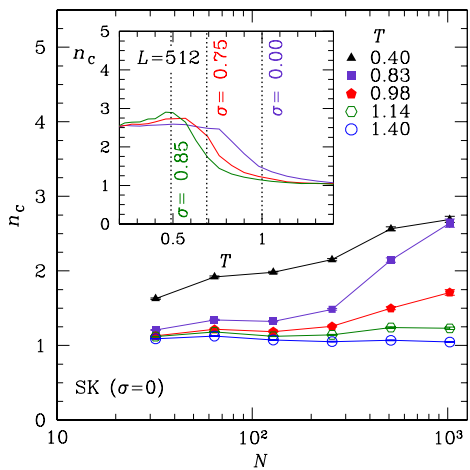}

\vspace*{-1.3cm}

\caption{(Color online)
Number of phase-space components $n_{\rm C}$ in the SK model as a
function of system size $N$ for different temperatures. For $T \lesssim
T_c$ (full symbols) the number of components grows considerably for
increasing system size $L$, whereas for $T \gtrsim T_c$ (open symbols)
the number of components remains approximately constant as a function
of $L$. The inset shows the number of components as a function of
temperature $T$ for $L = 512$ for different exponents $\sigma$. The
data for all $\sigma$ are qualitatively similar: for $T \sim T_c$
(dotted lines) the number of components is much larger than at $T
\gg T_c$.
}
\label{fig:clusters}
\end{figure}

In Fig.~\ref{fig:clusters} the number of components $n_{\rm C}$ is
shown for the SK model as function of $L$ for different $T$. Below
$T_c$ $n_{\rm C}$ increases with $L$, while for larger $T$ it
decreases. Other values of $\sigma$ show a similar behavior (not
shown).  Interestingly, $n_{\rm C}$ is largest in the spin-glass
phase and close to $T_c$ (inset to Fig.~\ref{fig:clusters}).
The reason is probably that at higher $T$ more energy landscape
valleys are accessible, including those who have high-lying minima,
still separated by energy barriers rarely overcome. For even higher
temperatures even more states are highly populated, leading to one
big component in the energy landscape. The exact peak position shifts
slightly with $\sigma$.

\paragraph*{Summary and discussion.---}
\label{sec:conclusions}

We have studied numerically the low-temperature configuration landscape
of a 1D long-range spin glass with power-law interactions characterized
by an exponent $\sigma$.  By using a hierarchical clustering method
and analyzing the resulting distance matrices we have studied the UM
properties, as well as counted the phase space components.  For this
purpose we have introduced a novel way to quantify ultrametricity
and we have extended a method to count components by analyzing the
distance matrix structure.  We observe that for $\sigma$ values
spanning the infinite-range SK universality class ($\sigma = 0$)
to the non-mean-field universality class ($\sigma > 2/3$) an UM
organization and a complex clustered landscape seem to emerge for
the system sizes studied. To check if these results persist at larger
length scales, it would be of interest to study even larger systems
\cite{leuzzi:08-ea}. This important since the crossover to any putative
UM behavior presumably might depend on the system size.

\begin{acknowledgments} 

We thank E.~Domany, G.~Hed, T.~J\"org, F.~Krzakala and A.~P.~Young
for the useful discussions, as well as T.~J\"org for providing
test data for the Migdal-Kadanoff spin glasses. We especially thank
W.~Radenbach for his participation in an initial stage of the project.
The simulations have been performed on the ETH Z\"urich clusters.
H.G.K.~acknowledges support from the Swiss National Science Foundation
under Grant No.~PP002-114713.

\end{acknowledgments}

\vspace{-0.5cm} \bibliography{refs,comments}

\begin{thebibliography}{27}
\expandafter\ifx\csname natexlab\endcsname\relax\def\natexlab#1{#1}\fi
\expandafter\ifx\csname bibnamefont\endcsname\relax
  \def\bibnamefont#1{#1}\fi
\expandafter\ifx\csname bibfnamefont\endcsname\relax
  \def\bibfnamefont#1{#1}\fi
\expandafter\ifx\csname citenamefont\endcsname\relax
  \def\citenamefont#1{#1}\fi
\expandafter\ifx\csname url\endcsname\relax
  \def\url#1{\texttt{#1}}\fi
\expandafter\ifx\csname urlprefix\endcsname\relax\def\urlprefix{URL }\fi
\providecommand{\bibinfo}[2]{#2}
\providecommand{\eprint}[2][]{\url{#2}}

\bibitem[{\citenamefont{Rammal et~al.}(1986)\citenamefont{Rammal, Toulouse, and
  Virasoro}}]{rammal:86}
\bibinfo{author}{\bibfnamefont{R.}~\bibnamefont{Rammal}},
  \bibinfo{author}{\bibfnamefont{G.}~\bibnamefont{Toulouse}}, \bibnamefont{and}
  \bibinfo{author}{\bibfnamefont{M.~A.} \bibnamefont{Virasoro}},
  \bibinfo{journal}{Rev. Mod. Phys.} \textbf{\bibinfo{volume}{58}},
  \bibinfo{pages}{765} (\bibinfo{year}{1986}).

\bibitem[{\citenamefont{Parisi}(1979)}]{parisi:79}
\bibinfo{author}{\bibfnamefont{G.}~\bibnamefont{Parisi}},
  \bibinfo{journal}{Phys. Rev. Lett.} \textbf{\bibinfo{volume}{43}},
  \bibinfo{pages}{1754} (\bibinfo{year}{1979}).

\bibitem[{\citenamefont{Binder and Young}(1986)}]{binder:86}
\bibinfo{author}{\bibfnamefont{K.}~\bibnamefont{Binder}} \bibnamefont{and}
  \bibinfo{author}{\bibfnamefont{A.~P.} \bibnamefont{Young}},
  \bibinfo{journal}{Rev. Mod. Phys.} \textbf{\bibinfo{volume}{58}},
  \bibinfo{pages}{801} (\bibinfo{year}{1986}).

\bibitem[{\citenamefont{{M{\' e}zard}~{\em et al.}}(1984)}]{mezard:84-ea}
\bibinfo{author}{\bibfnamefont{M.}~\bibnamefont{{M{\' e}zard}~{\em et al.}}},
  \bibinfo{journal}{Phys. Rev. Lett.} \textbf{\bibinfo{volume}{52}},
  \bibinfo{pages}{1156} (\bibinfo{year}{1984}).

\bibitem[{\citenamefont{Sherrington and Kirkpatrick}(1975)}]{sherrington:75}
\bibinfo{author}{\bibfnamefont{D.}~\bibnamefont{Sherrington}} \bibnamefont{and}
  \bibinfo{author}{\bibfnamefont{S.}~\bibnamefont{Kirkpatrick}},
  \bibinfo{journal}{Phys. Rev. Lett.} \textbf{\bibinfo{volume}{35}},
  \bibinfo{pages}{1792} (\bibinfo{year}{1975}).

\bibitem[{\citenamefont{M\'ezard et~al.}(1987)\citenamefont{M\'ezard, Parisi,
  and Virasoro}}]{mezard:87}
\bibinfo{author}{\bibfnamefont{M.}~\bibnamefont{M\'ezard}},
  \bibinfo{author}{\bibfnamefont{G.}~\bibnamefont{Parisi}}, \bibnamefont{and}
  \bibinfo{author}{\bibfnamefont{M.~A.} \bibnamefont{Virasoro}},
  \emph{\bibinfo{title}{Spin Glass Theory and Beyond}}
  (\bibinfo{publisher}{World Scientific}, \bibinfo{address}{Singapore},
  \bibinfo{year}{1987}).

\bibitem[{\citenamefont{Bray and Moore}(1986)}]{bray:86}
\bibinfo{author}{\bibfnamefont{A.~J.} \bibnamefont{Bray}} \bibnamefont{and}
  \bibinfo{author}{\bibfnamefont{M.~A.} \bibnamefont{Moore}}, in
  \emph{\bibinfo{booktitle}{Heidelberg Colloquium on Glassy Dynamics and
  Optimization}}, edited by
  \bibinfo{editor}{\bibfnamefont{L.}~\bibnamefont{Van~Hemmen}}
  \bibnamefont{and}
  \bibinfo{editor}{\bibfnamefont{I.}~\bibnamefont{Morgenstern}}
  (\bibinfo{publisher}{Springer}, \bibinfo{address}{New York},
  \bibinfo{year}{1986}), p. \bibinfo{pages}{121}.

\bibitem[{\citenamefont{Fisher and Huse}(1986)}]{fisher:86}
\bibinfo{author}{\bibfnamefont{D.~S.} \bibnamefont{Fisher}} \bibnamefont{and}
  \bibinfo{author}{\bibfnamefont{D.~A.} \bibnamefont{Huse}},
  \bibinfo{journal}{Phys. Rev. Lett.} \textbf{\bibinfo{volume}{56}},
  \bibinfo{pages}{1601} (\bibinfo{year}{1986}).

\bibitem[{\citenamefont{Krzakala and Martin}(2000)}]{krzakala:00}
\bibinfo{author}{\bibfnamefont{F.}~\bibnamefont{Krzakala}} \bibnamefont{and}
  \bibinfo{author}{\bibfnamefont{O.~C.} \bibnamefont{Martin}},
  \bibinfo{journal}{Phys. Rev. Lett.} \textbf{\bibinfo{volume}{85}},
  \bibinfo{pages}{3013} (\bibinfo{year}{2000}).

\bibitem[{\citenamefont{Palassini and Young}(2000)}]{palassini:00}
\bibinfo{author}{\bibfnamefont{M.}~\bibnamefont{Palassini}} \bibnamefont{and}
  \bibinfo{author}{\bibfnamefont{A.~P.} \bibnamefont{Young}},
  \bibinfo{journal}{Phys. Rev. Lett.} \textbf{\bibinfo{volume}{85}},
  \bibinfo{pages}{3017} (\bibinfo{year}{2000}).

\bibitem[{\citenamefont{Hed et~al.}(2004)\citenamefont{Hed, Young, and
  Domany}}]{hed:03}
\bibinfo{author}{\bibfnamefont{G.}~\bibnamefont{Hed}},
  \bibinfo{author}{\bibfnamefont{A.~P.} \bibnamefont{Young}}, \bibnamefont{and}
  \bibinfo{author}{\bibfnamefont{E.}~\bibnamefont{Domany}},
  \bibinfo{journal}{Phys. Rev. Lett.} \textbf{\bibinfo{volume}{92}},
  \bibinfo{pages}{157201} (\bibinfo{year}{2004}).

\bibitem[{\citenamefont{{Franz} and {Ricci-Tersenghi}}(2000)}]{franz:00}
\bibinfo{author}{\bibfnamefont{S.}~\bibnamefont{{Franz}}} \bibnamefont{and}
  \bibinfo{author}{\bibfnamefont{F.}~\bibnamefont{{Ricci-Tersenghi}}},
  \bibinfo{journal}{Phys. Rev. E} \textbf{\bibinfo{volume}{61}},
  \bibinfo{pages}{1121} (\bibinfo{year}{2000}).

\bibitem[{\citenamefont{{J{\"o}rg} and {Krzakala}}(2008)}]{joerg:08c}
\bibinfo{author}{\bibfnamefont{T.}~\bibnamefont{{J{\"o}rg}}} \bibnamefont{and}
  \bibinfo{author}{\bibfnamefont{F.}~\bibnamefont{{Krzakala}}},
  \bibinfo{journal}{Phys. Rev. Lett.} \textbf{\bibinfo{volume}{100}},
  \bibinfo{pages}{159701} (\bibinfo{year}{2008}).

\bibitem[{\citenamefont{{Contucci {\em et al.}}}(2007)}]{contucci:07-ea}
\bibinfo{author}{\bibfnamefont{P.}~\bibnamefont{{Contucci {\em et al.}}}},
  \bibinfo{journal}{Phys. Rev. Lett.} \textbf{\bibinfo{volume}{99}},
  \bibinfo{pages}{057206} (\bibinfo{year}{2007}).

\bibitem[{\citenamefont{{Contucci {\em et al.}}}(2008)}]{contucci:08-ea}
\bibinfo{author}{\bibfnamefont{P.}~\bibnamefont{{Contucci {\em et al.}}}},
  \bibinfo{journal}{Phys. Rev. Lett.} \textbf{\bibinfo{volume}{100}},
  \bibinfo{pages}{159702} (\bibinfo{year}{2008}).

\bibitem[{\citenamefont{{Kotliar} et~al.}(1983)\citenamefont{{Kotliar},
  {Anderson}, and {Stein}}}]{kotliar:83}
\bibinfo{author}{\bibfnamefont{G.}~\bibnamefont{{Kotliar}}},
  \bibinfo{author}{\bibfnamefont{P.~W.} \bibnamefont{{Anderson}}},
  \bibnamefont{and} \bibinfo{author}{\bibfnamefont{D.~L.}
  \bibnamefont{{Stein}}}, \bibinfo{journal}{Phys. Rev. B}
  \textbf{\bibinfo{volume}{27}}, \bibinfo{pages}{R602} (\bibinfo{year}{1983}).

\bibitem[{\citenamefont{Katzgraber and Young}(2003)}]{katzgraber:03}
\bibinfo{author}{\bibfnamefont{H.~G.} \bibnamefont{Katzgraber}}
  \bibnamefont{and} \bibinfo{author}{\bibfnamefont{A.~P.} \bibnamefont{Young}},
  \bibinfo{journal}{Phys. Rev. B} \textbf{\bibinfo{volume}{67}},
  \bibinfo{pages}{134410} (\bibinfo{year}{2003}).

\bibitem[{\citenamefont{Katzgraber and {Young}}(2005)}]{katzgraber:05c}
\bibinfo{author}{\bibfnamefont{H.~G.} \bibnamefont{Katzgraber}}
  \bibnamefont{and} \bibinfo{author}{\bibfnamefont{A.~P.}
  \bibnamefont{{Young}}}, \bibinfo{journal}{Phys. Rev. B}
  \textbf{\bibinfo{volume}{72}}, \bibinfo{pages}{184416}
  (\bibinfo{year}{2005}).

\bibitem[{\citenamefont{Hukushima and Nemoto}(1996)}]{hukushima:96}
\bibinfo{author}{\bibfnamefont{K.}~\bibnamefont{Hukushima}} \bibnamefont{and}
  \bibinfo{author}{\bibfnamefont{K.}~\bibnamefont{Nemoto}},
  \bibinfo{journal}{J. Phys. Soc. Jpn.} \textbf{\bibinfo{volume}{65}},
  \bibinfo{pages}{1604} (\bibinfo{year}{1996}).

\bibitem[{\citenamefont{Jain and Dubes}(1988)}]{jain:88}
\bibinfo{author}{\bibfnamefont{A.~K.} \bibnamefont{Jain}} \bibnamefont{and}
  \bibinfo{author}{\bibfnamefont{R.~C.} \bibnamefont{Dubes}},
  \emph{\bibinfo{title}{Algorithms for Clustering Data}}
  (\bibinfo{publisher}{Prentice-Hall}, \bibinfo{address}{Englewood Cliffs,
  USA}, \bibinfo{year}{1988}).

\bibitem[{rem()}]{remark:tree}
\bibinfo{note}{According to Ref.~\cite{hed:03}, from each subtree $T_{1a}$,
  $T_{1b}$ and $T_2$ one state is selected randomly. The root has two subtrees
  $T_1$ and $T_2$, $T_1$ being the larger one. $T_{1a}$ and $T_{2a}$ are the
  two subtrees of $T_1$.}

\bibitem[{com({\natexlab{a}})}]{comment:eytan}
\bibinfo{note}{Note that for the definition of $K$ in Ref.~\cite{hed:03} $P(K)
  \to \delta(0)$ for $L \to \infty$ also when the system is {\it trivially}
  ultrametric.}

\bibitem[{com({\natexlab{b}})}]{comment:thomas}
\bibinfo{note}{A.~K.~Hartmann, T.~J\"org, H.~G.~Katzgraber, and F.~Krzakala, in
  preparation}.

\bibitem[{\citenamefont{Kelley~{\em et al.}}(1996)}]{kelley:96-ea}
\bibinfo{author}{\bibfnamefont{L.~A.} \bibnamefont{Kelley~{\em et al.}}},
  \bibinfo{journal}{Prot. Engin.} \textbf{\bibinfo{volume}{9}},
  \bibinfo{pages}{1063} (\bibinfo{year}{1996}).

\bibitem[{com({\natexlab{c}})}]{comment:log}
\bibinfo{note}{The value $d_{\alpha\beta}=1$, i.e., two equal configurations,
  does occur with exponentially-small probability for large systems but was
  never observed.}

\bibitem[{com({\natexlab{d}})}]{comment:clusters}
\bibinfo{note}{If one required an average value of exactly $1$ component(s),
  the sensitivity would depend on the number of samples.}

\bibitem[{\citenamefont{{Leuzzi}~{\em et al.}}(2008)}]{leuzzi:08-ea}
\bibinfo{author}{\bibfnamefont{L.}~\bibnamefont{{Leuzzi}~{\em et al.}}},
  \bibinfo{journal}{Phys. Rev. Lett.} \textbf{\bibinfo{volume}{101}},
  \bibinfo{pages}{107203} (\bibinfo{year}{2008}).

\end{thebibliography}

\end{document}